# Probabilities of spurious connections in gene networks: Application to expression time series

David R. Bickel*

*Office of Biostatistics and Bioinformatics*
*Medical College of Georgia*
*Augusta, GA 30912-4900*

**Motivation:** The reconstruction of gene networks from gene expression microarrays is gaining popularity as methods improve and as more data become available. The reliability of such networks could be judged by the probability that a connection between genes is spurious, resulting from chance fluctuations rather than from a true biological relationship.
**Results:** Unlike the false discovery rate and positive false discovery rate, the decisive false discovery rate (dFDR) is exactly equal to a conditional probability without assuming independence or the randomness of hypothesis truth values. This property is useful not only in the common application to the detection of differential gene expression, but also in determining the probability of a spurious connection in a reconstructed gene network. Estimators of the dFDR can estimate each of three probabilities:
1. The probability that two genes that appear to be associated with each other lack such association.
2. The probability that a time ordering observed for two associated genes is misleading.
3. The probability that a time ordering observed for two genes is misleading, either because they are not associated or because they are associated without a lag in time. The first probability applies to both static and dynamic gene networks, and the other two only apply to dynamic gene networks.
**Availability:** Cross-platform software for network reconstruction, probability estimation, and plotting is free from http://www.davidbickel.com as R functions and a Java application.
**Supplementary information:** Color figures are available from http://www.davidbickel.com.
**Contact:** bickel@prueba.info

*Address after March 31, 2004: Pioneer, Bioinformatics and Discovery Research, 7250 NW 62nd Ave., P. O. Box 552, Johnston, IA 50131-0552

# 1 Introduction

Variations of the false discovery rate of Benjamini and Hochberg (1995) have been successfully applied to the problem of detecting differential gene expression between two or more groups on the basis of microarray data (Efron *et al.* 2001; Efron and Tibshirani 2002; Müller *et al.* 2002; Pepe *et al.* 2003; Storey 2002, 2003; Bickel 2004a, b). In this context, a *discovery* of differential expression is the rejection of the null hypothesis that a gene is not differentially expressed, and such a discovery is false if there really is no differential expression, i.e., if the rejected null hypothesis is true. Conversely, a nondiscovery of differential expression is the nonrejection of a null hypothesis, so a nondiscovery is false if there is differential expression, i.e., if the nonrejected null hypothesis is false. False discovery rate methods not only tend to have much lower false nondiscovery rates than methods of controlling family-wise error rates, but they have a simpler interpretation: depending on which mathematical definition of the false discovery rate is used, it is either exactly equal to a probability under general conditions (Efron and Tibshirani 2002; Fernando *et al.* 2004), or it is approximately equal to a probability under more restrictive conditions (Efron *et al.* 2001; Storey 2002, 2003). The probability in question is the probability that a gene considered differentially expressed is not really differentially expressed. More generally, it is the probability that a discovery is false, which, by definition, is the probability that a rejected null hypothesis is true. This property of the false discovery rate will be used to answer the question, "What is the probability that a given relationship in a gene network reconstructed from gene expression data is spurious?" Thus, it will be seen that the methodology of false discovery rates not only aids in the detection of differential expression, but also in another important use of microarray technol-ogy, that of reverse-engineering regulatory networks of genes.

Many models of gene networks have been applied to the analysis of microarray data (De Jong 2002), and the following false discovery rate methods are suitable for all gene networks that specify definite relationships between genes, including any network that can be represented as an undirected or directed graph. Those methods will be illustrated with a network that can be seen as a dynamic generalization of networks constructed by considering two genes to be connected if the absolute value of the correlation coefficient between their expression values is sufficiently high. There are several variations of such networks (Butte *et al.*. 2000; Rho, Jeong, and Kahng 2003), all of which fall in the general class of *spatial networks* (Herrmann, Berthélemy, and Provero 2003), an example of a type of network that could benefit from the proposed methods.

Let $x_1(t)$ and $x_2(t)$ represent the expression values of the first and second gene at time $t$, respectively. (There are many possible definitions of *expression value*; some possibilities are given below, but here it is just assumed that the expression value tends to increase with the amount of mRNA.) Without loss of generality, define the *coexpression value* to be the absolute value of the linear correlation coefficient. Then, if $x_1(t)$ and $x_2(t)$ are weakly stationary ($t \in \mathbb{R}$), the coexpression value between the two genes at a lag time of $\tau$ is

$$|r_{1,2}(\tau)| = |r(x_1(0), x_2(\tau))| = |r(x_1(-\tau), x_2(0))|. \qquad (1)$$

If $|r_{1,2}(\tau)| > 0$, then the two genes are said to be *coregulated*. (Alternately, they are coregulated only if $|r_{1,2}(\tau)|$ is greater than or equal to the minimum coregulation that is biologically meaningful in the sense of Bickel (2004a).) Let $\tau_{\text{optimal}}$ be the time lag at which the coregulation is maximal, i.e., $\tau_{\text{optimal}}$ satisfies $|r_{1,2}(\tau_{\text{optimal}})| = \max_\tau |r_{1,2}(\tau)|$. Then if the two genes are coregulated, gene 2 is said to *lag behind* gene 1 by $\tau_{\text{optimal}}$ if $\tau_{\text{optimal}} > 0$, gene 1 is said to

lag behind gene 2 by $|\tau_{\text{optimal}}|$ if $\tau_{\text{optimal}} < 0$, and the genes are said to be *contemporaneous* in expression if $\tau_{\text{optimal}} = 0$. For a finite number of time points $n$, $t \in \{0, \Delta t, 2\Delta t, \ldots, (n-1)\Delta t\}$, $\Delta t > 0$, and the corresponding coexpression value is

$$|\hat{r}_{1,2}(\tau)| = \begin{cases} |r((x_{1,1+|\tau|/(\Delta t)}, x_{1,2+|\tau|/(\Delta t)}, \ldots, x_{1,n}), (x_{2,1}, x_{2,2}, \ldots, x_{2,n-|\tau|/(\Delta t)}))| & \tau \in \{-(n-1)\Delta t, \ldots, -2\Delta t, -\Delta t\} \\ |r((x_{1,1}, x_{1,2}, \ldots, x_{1,n}), (x_{2,1}, x_{2,2}, \ldots, x_{2,n}))| & \tau = 0 \\ |r((x_{1,1}, x_{1,2}, \ldots, x_{1,n-\tau/(\Delta t)}), (x_{2,1+\tau/(\Delta t)}, x_{2,2+\tau/(\Delta t)}, \ldots, x_{2,n}))| & \tau \in \{\Delta t, 2\Delta t, \ldots, (n-1)\Delta t\} \end{cases}, \quad (2)$$

with $x_{l,j} \equiv x_l((j-1)\Delta t)$. Here, values are truncated from both vectors when $\tau \neq 0$, but other boundary conditions are possible, and a boundary condition that only causes truncation from one vector is presented below. Exactly how boundary expression values are handled has little effect on the coexpression value when $|\tau| \lesssim n\Delta t/4$.

The methods of the next section construct a genetic network, from which one can make inferences about gene coexpression (1) based on expression levels obtained for thousands of genes from microarrays at fixed time intervals. The first method determines which combinations of gene pairs and lag times have finite series coexpression values (2) that meet or exceed a given threshold. The next method estimates the probability that any such combination is spurious in that its coexpression value (2) was as high as observed due to chance, rather than due to a nonzero true coexpression value (1). Given the set of gene-pair/lag-time combinations, the third method determines which of the nonzero lag times differ significantly from zero, and then estimates the probability that a significant nonzero lag time is a finite-size effect, i.e., that $\max_\tau |r_{1,2}(\tau)| = |r_{1,2}(0)|$ when genes 1 and 2 have a statistically significant nonzero lag time. Methods to describe the tails of the connectivity distributions follow. Joint inference based on the probabilities of the second and third methods is also described. Lastly, yeast cell cycle data are analyzed for the purpose of illustration.

## 2 New methods

### 2.1 Finding gene networks

Given *m* genes, each with expression values measured at *n* equally-spaced time points, all of the expression values can be represented by an $m \times n$ matrix, **X**. (This representation and method also applies to the more general cases of *n* experiments, not necessarily with a time-series design, and of *n* laboratory animals, human patients, or other individuals.) Denote by $x_{i,j}$ the entry in the *i*th row and *j*th column of **X**, i.e., $x_{i,j}$ is the expression value of the *i*th gene and the *j*th time point, individual, or experiment. The term "expression value" applies to any measure of a relative amount of gene expression, including logarithms or ranks of estimated mRNA levels or ratios. The ranks of the estimated mRNA levels within each gene are particularly useful since they are insensitive to outliers and to distributional differences. For the *i*th gene, the rank for the *j*th time point, experiment, or individual is the rank of the corresponding estimated mRNA level or ratio among all the measurements for that gene, so that the rank is equal to 1 for the smallest, 2 for the second smallest, *k* for the *k*th smallest, and *n* for the largest estimated mRNA level or ratio for the *i*th gene. This is implemented in software by applying the rank transform independently to each row of an $m \times n$ matrix of estimated mRNA levels or ratios, resulting in **X**, an $m \times n$ matrix of expression ranks. Two genes are considered to be *coregulated* if their row vectors in **X** satisfy a coregulation criterion that is based on a measure of similarity, such as correlation, or a measure of dissimilarity, such as Euclidean distance. In this paper, the coregulation criterion *C* is satisfied for two rows if the absolute value of their correlation coefficient is at least $r_0$, where $r_0$ is some specified value between 0 and 1. Higher values of $r_0$ correspond to less inclusive (more stringent) criteria. For the sake of conciseness, the term "coregulation value" will include all measures of association, including the absolute value of the correlation and distance-based similarities.

The following algorithm generates a network of genes from **X**, given a coregulation criterion $C$ and, for time-series data, a maximum lag time $\tau_{max}$, a nonnegative integer multiple of $\Delta t$, the time between two consecutive measurements. Since correlations become unreliable as their lag times approach the length of the time series, it is recommended that $\tau_{max}$ be not much more than a fourth of the duration of the experiment. $C$ may be chosen to be as inclusive as possible, subject to limitations of computer resources, since it can easily be made more stringent after initial analysis, as seen below.

**Step 1.** If the columns of **X** correspond to equally-spaced points in time, proceed to Step 2. Otherwise, set **Y** equal to **X** (**Y** ← **X**, i.e., let **Y** = **X**), $m' \leftarrow m$, and proceed to Step 4.

**Step 2.** Create time-shifted matrices by translating the data matrix to the left and to the right, as follows:
For $\tau = -\tau_{max}, -\tau_{max} + \Delta t, -\tau_{max} + 2\Delta t, \ldots, -2\Delta t,$
$-\Delta t, 0, \Delta t, 2\Delta t, \ldots, \tau_{max} - 2\Delta t, \tau_{max} - \Delta t, \tau_{max},$
let $\mathbf{X}(\tau)$ be the $m \times n$ matrix created by shifting **X** by $|\tau/(\Delta t)|$ columns to the right if $\tau \geq 0$ or to the left if $\tau < 0$, with the values of $|\tau/(\Delta t)|$ columns dropped from one side of the matrix and with padding that repeats the closest column with expression values on the other side of the matrix. Thus, $\mathbf{X}(\tau)$ is a matrix of $m$ row vectors, the $i$th of which is $(x_{i,1}(\tau), x_{i,2}(\tau), \ldots, x_{i,n}(\tau))$, where $x_{i,j}(\tau)$ is the expression level of the $i$th gene and the $(j - \tau/(\Delta t))$th time point unless $j - \tau/(\Delta t) < 1$ or $j - \tau/(\Delta t) > n$. For example, if $\tau = 2\Delta t$, then
$\forall_{i \in \{1,2,\ldots,m\}} x_{i,1}(2\Delta t) = x_{i,2}(2\Delta t) = x_{i,1}$ and $\forall_{i \in \{1,2,\ldots,m\}; j \in \{3,4,\ldots,n\}} x_{i,j}(2\Delta t) = x_{i,j-2}$. (It follows that $\mathbf{X}(0) = \mathbf{X}$.)

**Step 3.** The matrix **Y** is created by combining the rows of all of the matrices of Step 2. $m' \leftarrow (2|\tau_{max}/(\Delta t)|+1)m$, so **Y** is an $m' \times n$ matrix.

**Step 4.** Compute all possible row-row coregulation values for **Y**, recording which pairs of rows that have sufficient coregulation to satisfy $C$. [This computationally intensive step was implemented in an R operating system call to a Java application, whereas all of the other computations herein were performed solely by functions written in R (R Development Core Team, 2003).]

**Step 5.** Interpret the results of Step 4 in terms of which genes are coregulated, and at which lag times they are coregulated. This is accomplished by mapping the rows of **Y** back to the rows of **X** and values of $\tau$ that were used to construct **Y**, noting the relationships between rows of **Y** found in Step 4. Thus, each pair of rows of **Y** that has sufficient coregulation corresponds to a pair of genes and a pair of lag times. These pairs of rows of **Y** will be referred to as the *coregulated row pairs*.

**Step 6.** Using the results of Step 5, for all coregulated row pairs that satisfy *C* for different values of $\tau$, record which gene of each pair lags behind the other gene of its pair, the amount of lag time between the two, and the coregulation value. For example, if a coregulated row pair has two genes that satisfy *C* when the row of the first gene in **Y** was from $\mathbf{X}(\tau_1)$ and the row of the second gene in **Y** was from $\mathbf{X}(\tau_2)$, with $\tau_1 < \tau_2$, then the second gene lags behind the first gene by $|\tau_1 - \tau_2|$. When more than one coregulated row pair give the same gene and relative lag time, their highest coregulation value is taken as representative of the relationship between the two genes at that relative lag time. (The highest value instead of an average value avoids a bias that would be present in an average value since the coregulated row pairs in Step 4 were chosen for having high coregulation values, leaving out row pairs with the same gene and lag information that did not satisfy *C*.) (For the following application to yeast data, the maximum absolute value of the correlation coefficient is the highest coregulation value.) Then, each pair of genes and relative lag time with a sufficiently high similarity can be represented by a *gene-gene relationship* consisting of a *leading gene*, a *lagging gene*, a *relative lag time*, a *coregulation value*, and the *direction of association* (positive or negative). In the yeast example, the coregulation value and the direction of the association are the magnitude and sign of the correlation coefficient, respectively. By contrast, distance-based measures of coregulation typically only account for positive association, although there are exceptions, e.g., Bickel (2003). When the relative lag time is zero, the "leading gene" and "lagging gene" are interchangeable. For gene-gene relationships that have the same pair of genes, only retain the relationships with coregulation values that are local maxima with respect to lag time, in order to eliminate relationships that have high coregulations only because they are close in time to relationships with higher coregulations. All *K* of the remaining gene-gene relationships constitute $\mathcal{N}$, the *network* of genes.

The open source code exemplifies a detailed implementation of this algorithm. The network of genes thereby determined may be represented by a directed graph if there is time information, as shown below. Otherwise, it may be represented as an undirected graph. In either case, the edges of the graph connect nodes that represent coregulated genes, with each edge representing a gene-gene relationship.

Instead of this algorithm, Kellam *et al.* (2002) used an evolutionary program to find the gene-gene relationships that satisfied the criterion for the correlation coefficient without the rank transform. Thus, the same measure of similarity may be used with different transforms and algorithms. Non-rank transforms may improve statistical power if the data fit a known model well, in which case the permutation test could possibly be replaced by a parametric test. Which algorithm is best for network reconstruction also depends on the data, but the suggested algorithm is widely applicable and conservative, even when not ideal.

**2.2 Probabilities of spurious connections**

Some of the gene-gene relationships in the network are *false discoveries* in that they are finite-$n$ effects that would not appear in a large number of repeated experiments. To get an idea of how many discoveries are false, there are many ways to quantify a "false discovery rate"; the simple *decisive false discovery rate* (dFDR) of Bickel (2004a, b) will be used here. The estimated dFDR is zero if there are no gene-gene relationships. [By contrast, the closely related estimated proportion of false positives (PFP) of Fernando *et al.* (2004) would be undefined.] When there are gene-gene relationships in the network, the estimated dFDR is equal to the estimated proportion of false positives, the ratio of $R_0$, the number of gene-gene relationships that would occur under a null hypothesis, to $R$, the number of gene-gene relationships in the network obtained from the data. A meaningful null hypothesis is that the rows and columns of **X** are independent. In this case, $R_0$ could be the number of gene-gene relationships found by applying Steps 1–6 to an $m \times n$ matrix formed by randomly permuting the values within each of the rows of **X**. The random effects of the permutations can be reduced by forming $B'$ independent matrices by the same permutation method, in which case $R_0$ would be the mean number of gene-gene relationships found, i.e., the total number of gene-gene relationships found, divided by $B'$. The dFDR is estimated by $\min(1, R_0/R)$ if $R > 0$ or by 0 if $R = 0$, giving an idea of how reliable the gene-gene relationships are. Like the PFP, the dFDR can be viewed as the probability $p_{\text{spurious}}(C)$ that a given gene-gene relationship that satisfies $C$ is spurious (Bickel, 2004a, b). (The dFDR equals a conditional probability in the sense of Breiman (1992), even when the PFP is undefined.) Thus, a dFDR estimate close to 1 ($\hat{p}_{\text{spurious}}(C) \approx 1$) indicates that there is evidence for few or no real gene-gene relationships. This estimate is very conservative, and multiplying it by the estimated portion of true null hypotheses can reduce the number of false nondiscoveries (Storey 2002; 2003). As noted in the introduction, this dFDR methodology is not limited to the above algorithm. For example, parametric or nonparametric tests that assign a p-value to each gene-gene relationship would use the product of the significance level and the number of tests as $R_0$, using an established way to estimate a false discovery rate given a list of p-values (Genovese and Wasserman 2002; Efron and Tibshirani 2002; Storey 2003; Fernando *et al.* 2004; Bickel 2004a, b), as described in Section 2.3 in the context of discoveries of nonzero lag times.

Given a gene network found by the above algorithm, a gene network $\mathcal{N}'$ with a more stringent coregulation criterion $C'$ can be quickly found without repeating Steps 1–6. ($C'$ can be said to be *more stringent than C* only if every pair of vectors that satisfies $C'$ also satisfies $C$. For example, if the absolute value of the correlation is the coregulation value, then $C$ might require that the coregulation value is at least 0.9, whereas $C'$ might require that it is at least 0.99.) $\mathcal{N}'$ is the set of gene-gene relationships of $\mathcal{N}$ with coregulation values that satisfy $C'$. $\hat{p}_{\text{spurious}}(C')$ can then be computed in the same way as $\hat{p}_{\text{spurious}}(C)$.

## 2.3 Determining statistical significance of estimated lag times

While permuting the data of **X** provides $\hat{p}_{\text{spurious}}$, a good estimate of the probability that a gene-gene relationship is spurious due to possible independence, it does not provide $\hat{p}_{\text{contemporaneous}}$, an estimate of $p_{\text{contemporaneous}}$, the probability that a gene-gene relationship has a spuriously nonzero relative lag time, a positive relative lag time resulting from the finite size of the data set rather than from one of the genes really preceding the other in gene expression patterns. This probability is a dFDR that can be estimated by the ratio of the mean number of positive relative lag times that would occur by chance to the number of positive relative lag times in the network generated from the data. It can be based on a p-value for each positive relative lag time.

The p-value of the $k$th of $K$ relative lag times is the estimated probability that it is as great or greater than the relative lag time observed would occur if the genes had contemporaneously fluctuating expression values:

$$P_k = \frac{1}{B} \sum_{b=1}^{B} I(\tau_{k,b} \geq \tau_k), \tag{3}$$

where $I$ is an indicator function, $\tau_k$ is the relative lag time for the $k$th gene-gene relationship from **X**, and $\tau_{k,b}$ is the relative lag time of the $b$th of $B$ bootstrap samples of the $k$th gene-gene relationship, so that $I(\tau_{k,b} \geq \tau_k) = 1$ if $\tau_{k,b} \geq \tau_k$ and $I(\tau_{k,b} \geq \tau_k) = 0$ if $\tau_{k,b} < \tau_k$. (In the terminology of Efron and Tibshirani (1993), $P_k$ is the *achieved significance level*.) Each bootstrap sample for the $k$th gene-gene relationship is generated as follows when the expression values are ranks. (Some modification is needed for most other measures of expression value). Let $i(\text{lead}, k)$ be the index of the row in $\mathbf{X}$ of the leading gene, and let $i(\text{lag}, k)$ be that of the lagging gene. Then the base level of expression for the $j$th time point of those genes under the null model is

$$\overline{x}^*_{k,j} = \frac{1}{2} (x_{i(\text{lead},k),j} + x_{i(\text{lag},k),j}), \tag{4}$$

the $k$th set of residuals is

$$\mathcal{R}_k = \{x_{i(\text{lead},k),j} - \overline{x}^*_{k,j}\}_{j=1}^n \cup \{x_{i(\text{lag},k),j} - \overline{x}^*_{k,j}\}_{j=1}^n, \tag{5}$$

and the level of expression at the $j$th time point for the $l$th "gene" of the the $b$th bootstrap sample for the $k$th gene-gene relationship is

$$x^*_{k,b,l,j} = \overline{x}^*_{k,j} + \Delta x^*_{k,b,l,j}, \tag{6}$$

where $\Delta x^*_{k,b,l,j}$ is randomly selected with replacement from $\mathcal{R}_k$ and $l \in \{1, 2\}$. $\tau_{k,b}$ is the absolute value of the lag time that maximizes the coexpression value between the first and second bootstrap gene, that is, between the vectors $(x^*_{k,b,1,1}, x^*_{k,b,1,2}, \ldots, x^*_{k,b,1,n})$ and $(x^*_{k,b,2,1}, x^*_{k,b,2,2}, \ldots, x^*_{k,b,2,n})$. The maximum is taken over some set of lag times that includes 0 and $\tau_k$.

All relative lag times with p-values less than or equal to some significance level $\alpha$ are *statistically significant*. Because of the multiple comparisons issue, the statistical significance of a relative lag time does not provide evidence for a biological lag time, except for small $p_{\text{contemporaneous}}(C; \alpha)$, the probability that a relative lag time significant at level $\alpha$ corresponds to a true lag time of zero. That probability is estimated by

$$\hat{p}_{\text{contemporaneous}}(C; \alpha) = \frac{\alpha K}{\sum_{k=1}^{K} I(P_k \leq \alpha)}, \tag{7}$$

with the dependencies on *C* suppressed for notational simplicity.

The probability that either type of false discovery occurs for a significant relative lag time can be estimated by the sum of the two dFDR estimates: $\hat{p}_{\text{error|disjoint}}(C; \alpha) = \hat{p}_{\text{spurious}}(C) + \hat{p}_{\text{contemporaneous}}(C; \alpha)$, where $\hat{p}_{\text{spurious}}(C)$ is computed only for nonzero relative lag times. This may be too conservative in some cases since the two false discovery events are not mutually exclusive. A more reasonable estimate is based on the independence assumption:

$$\hat{p}_{\text{error|independent}}(C; \alpha) = \hat{p}_{\text{spurious}}(C) + \hat{p}_{\text{contemporaneous}}(C; \alpha) - \hat{p}_{\text{spurious}}(C)\, \hat{p}_{\text{contemporaneous}}(C; \alpha). \tag{8}$$

However, unless both dFDR estimates are very large, the two estimates differ only negligibly. For example, if $\hat{p}_{\text{spurious}}(C) = \hat{p}_{\text{contemporaneous}}(C; \alpha) = 5\,\%$, then $\hat{p}_{\text{error|disjoint}}(C; \alpha) = 10\,\%$ and $\hat{p}_{\text{error|independent}}(C; \alpha) = 9.75\,\%$.

That these are estimates, not known probabilities, must be kept in mind, especially when there are only a few time points. A careful simulation study of the mean square error of the probability estimates as a function of the number of genes and number of time points would be valuable, but no such study is planned at this time. As in other cases of false discovery rate estimation, reporting confidence intervals of estimates can prevent undue reliance on them (Bickel 2004c).

A directed graph of all gene-gene relationships with statistically significant relative lag times has genes as nodes and a directed edge for each gene-gene relationship, from the node of the leading gene to the node of the lagging gene, with the relative lag time and the direction of association ("+" or "−") as the edge label. Many of the directed edges may result from causative relationships between the genes, but a directed edge is insufficient to prove causation.

## 2.4 Scale-free networks

The extent to which a network is scale-free can be quantified by the fit of the tail of $f$, the probability density of the connectivity, to a power-law. The *connectivity* of a gene, denoted by $K$, is its number of connections, the number of genes with which it shares one or more gene-gene relationships. Agrawal (2002) found that networks constructed by a mutual nearest-neighbor algorithm satisfy

$$f(\tilde{K}) \propto \tilde{K}^\beta; \quad \tilde{K} \equiv (K+1)/(K_{\max}+1) \in (0, 1] \tag{9}$$

for $\tilde{K} \gtrsim 0.35$ with $\beta = -1$, where $K_{\max}$ is the highest connectivity in the network. Networks that satisfy equation (9) are considered scale-free or self-similar since $f(\tilde{K}_2) \approx (\tilde{K}_2/\tilde{K}_1)^\beta f(\tilde{K}_1)$ in the tails. The lack of a characteristic scale implies that the mean connectivity does not provide a good description of the network since it has many nodes with low connectivity and a few nodes with connectivity of one or more orders of magnitude higher.

Integrating (9) gives the cumulative probability function $F$, the probability that the number of connections is at least $K$:

$$F(\tilde{K}; a_1, a_2, \beta) = \int f(\tilde{K}) \, d\tilde{K} = \begin{array}{ll} a_1 - a_2 \ln(\tilde{K}); & \beta = -1 \\ a_1 - \frac{a_2}{1+\beta} \left( \tilde{K}^{(1+\beta)} - 1 \right); & \beta < -1 \end{array} \tag{10}$$

for sufficiently large $\tilde{K}$. Agrawal (2002) set $\beta = -1$ and used equation (10) to obtain least-squares estimates of the parameters $a_1$ and $a_2$ from empirical values of $F(\tilde{K})$. Nonlinear least-squares regression can be used to estimate $\beta$ as well as $a_1$ and $a_2$. Alternately, $\beta$ can be estimated by linear regression using the two-parameter model

$$F(\tilde{K}; a, \beta) = a\,\tilde{K}^{1+\beta}; \quad \beta < -1 \qquad (11)$$

since

$$\ln F(\tilde{K}; a, \beta) = \ln a + (1 + \beta) \ln \tilde{K} \qquad (12)$$

is the equation of a line.

## 3 Application to cell cycle data

Spellman *et al.* (1998) carried out an experiment to measure the variation of expression levels of yeast genes with the cell cycle. They noted that cell cycle regulation could keep order in cell division or conserve resources. Many genes both control and are controlled by the cell cycle, and most of the control is transcriptional (Spellman *et al.*, 1998). In the experiment, they applied a pheromone ($\alpha$-factor) to a yeast culture in order to stop the cells in the same cell age (G1, the first phase). Upon removing the pheromone, they sampled the culture and took a repeated sample every 7 minutes for 119 minutes, yielding a total of $n = 18$ samples. For each sample, a cDNA microarray was used to measure the expression values of 6178 genes. The methods described above were applied to the within-gene ranks of expression ratios of the $m = 4489$ genes that have no expression values missing in any of the microarrays. Applying the rank transform before computing the correlation coefficient in effect uses Spearman's correlation coefficient, which was informative in another expression time series in yeast (Bickel, 2003). The use of ranks minimizes the effects of outliers and obviates logarithmic or other monotonic transformation.

The above algorithm of network construction was used with $\tau_{\max} = 21$ minutes and coregulation thresholds 0.90, 0.91, 0.92, 0.93, 0.94, and 0.95. Fig. 1 summarizes all of the resulting gene-gene relationships, whereas Fig. 2 only summarizes those for which $\tau \neq 0$. Each plotted estimated probability that a relationship is false is $\hat{p}_{\text{spurious}}$, the dFDR estimated from $B' = 4$ independent permutations of all rows of the matrix.

Figs. 3 and 4 display the entire directed graph and partial directed graph of the network, respectively, for the coregulation threshold of 0.95, $\tau \neq 0$, and significance level $\alpha = 0.01$ for $B = 250$ iterations of the bootstrap procedure. Estimation of dFDRs yields $\hat{p}_{\text{spurious}}(|r| \geq 0.95) \approx 1.6\%$ and $\hat{p}_{\text{contemporaneous}}(|r| \geq 0.95; 0.01) \approx 2.9\%$, so the overall probability for error is approximately $\hat{p}_{\text{error|independent}}(|r| \geq 0.95; 0.01) \approx \hat{p}_{\text{error|disjoint}}(|r| \geq 0.95; 0.01) \approx 4.5\%$. Of the 37 genes in gene-gene relationships that have a coregulation of at least 0.95, a nonzero lag time, and a p-value of 0.01 or lower, four genes were not included in the set of 800 genes that Spellman *et al.* (1998) reported to be cell-cycle regulated; their open reading frame labels are YNL118C, YPR028W, YCL064C, and YGL139W. YNL118C has two gene-gene relationships satisfying the criteria, as seen in Fig. 4, but each of the other three genes only has one gene-gene relationship. The expression ranks of YNL118C and of the two genes connected to it were plotted in Fig. 5. As Fig. 4 indicates, YNL118C is 3 time units (21 minutes) ahead of the two lagging genes, which is also clear from part **b** of Fig. 5. YNL118C is known to prevent translation by allowing the production of active Dcp1p, the yeast mRNA decapping enzyme. While such posttranscriptional regulation would not in itself change the mRNA expression levels of the two lagging genes, YNL118C might reduce the abundance of transcription factors that regulate those genes. Some caution is needed here since the arrows in the directed graphs suggest causality without strictly implying it.

Using the distributions of connectivity at different coregulation thresholds, the overall structures of the networks reconstructed from the cell cycle data were also studied. Fig. 6 gives the total number of connections and the estimate of $\beta$ determined from fitting the three-parameter model of equation (10). The results for the coregulation threshold of 0.95 are not shown since its small number of relationships makes them unreliable for regression. Estimates are displayed for the three-parameter model since it fits the data much better than either of the two-parameter models, as exemplified by Fig. 7. Estimates of the scaling exponent are between $-3$ and $-2$.

## 4 Acknowledgements

## Figure Captions

(**a**) Number of relationships and (**b**) estimated probability that a relationship is false versus the coregulation threshold for all gene-gene relationships.

**Figure 1**

(**a**) Number of relationships and (**b**) estimated probability that a relationship is false versus the coregulation threshold for all gene-gene relationships with nonzero time lags.

**Figure 2**

The entire network for the coregulation threshold of 0.95.

### Figure 3

Selected parts of the network for the coregulation threshold of 0.95.

### Figure 4

Each unit of time on the horizontal axis corresponds to 7 minutes for (**a**) expression ranks of three genes and (**b**) expression ranks of the two lagging genes with the expression ranks of the leading gene reflected. The reflection is necessary to show the relationship between the genes at a lag time of 3 units (21 minutes) since they are negatively correlated. Color key: YNL118C is black, YJL201W is red, and YPR076W is green. The online supplementary information has color figures.

### Figure 5

(**a**) The number of connections and (**b**) the scaling exponent estimate, $\hat{\beta}$, versus the coregulation threshold for all gene-gene relationships.

### Figure 6

Cumulative probability distribution, the probability that the connectivity is greater than or equal to $K$, versus $K$ for coregulation thresholds of (**a**) 0.90 and (**b**) 0.93. The curves are least-squares fits, with equation (10) and $\beta = -1$ represented by black, equation (10) and $\beta$ estimated by the regression represented by blue, and equation (12) represented by red. The blue curve was fit using the nls function (R Development Core Team, 2003), with initial parameter values $\beta = -1.5$, $a_1 = 0.055$, and $a_2 = 0.6$ since those values of $a_1$ and $a_2$ were typical in Agrawal (2002) for $\beta = -1$. The online supplementary information has color figures.

### Figure 7

*D. R. Bickel*

# Figure 1

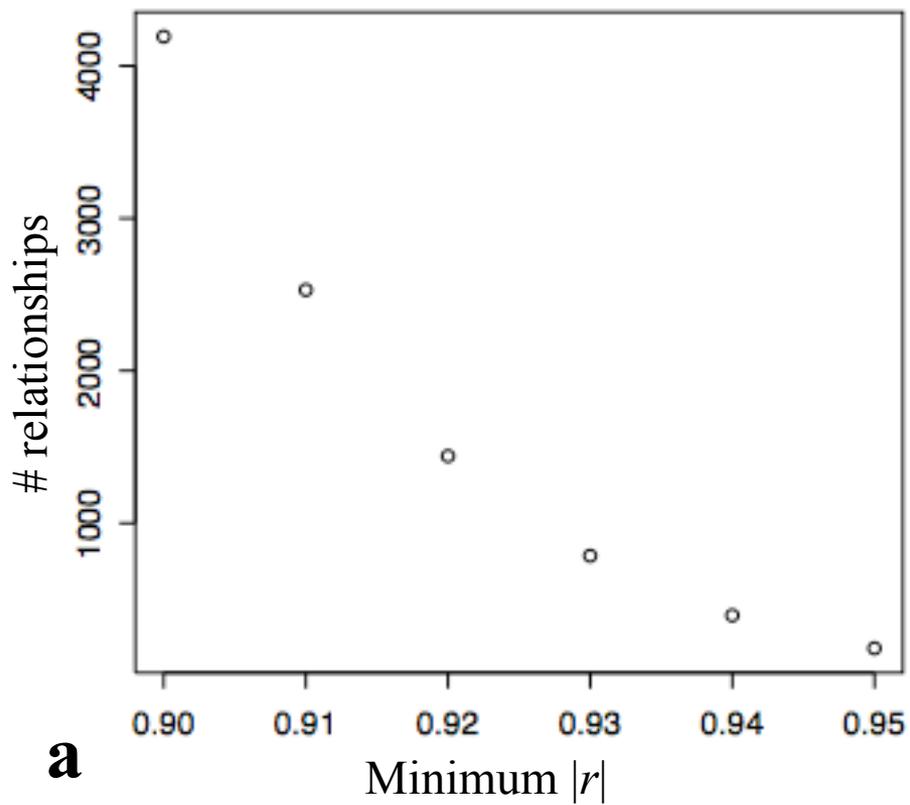 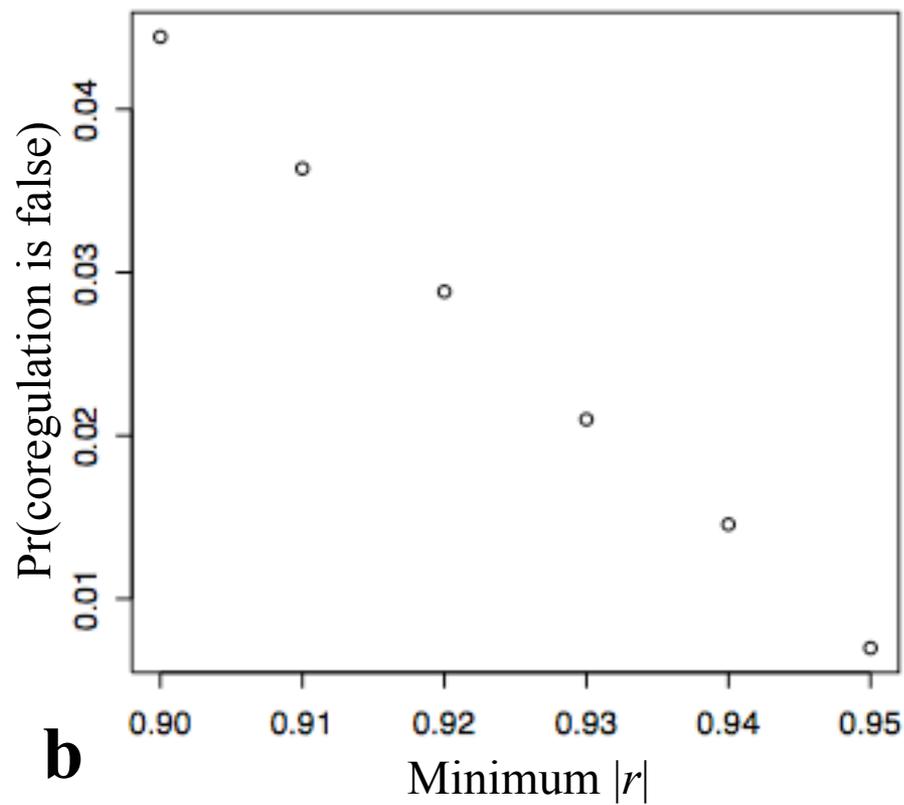

# Figure 2

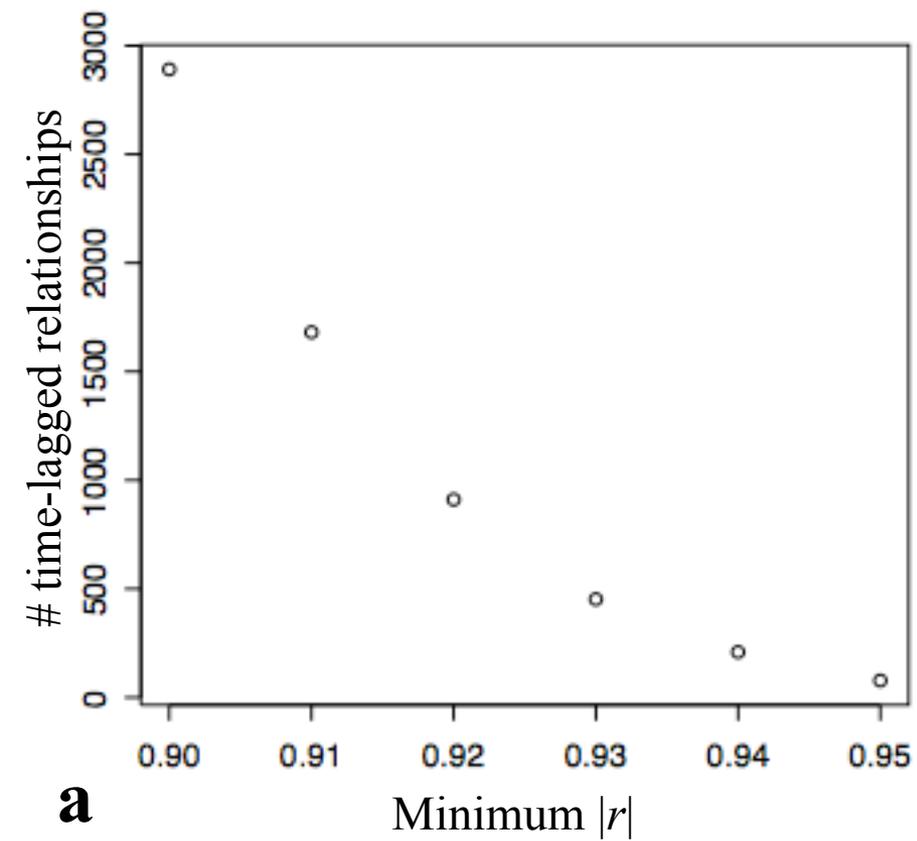 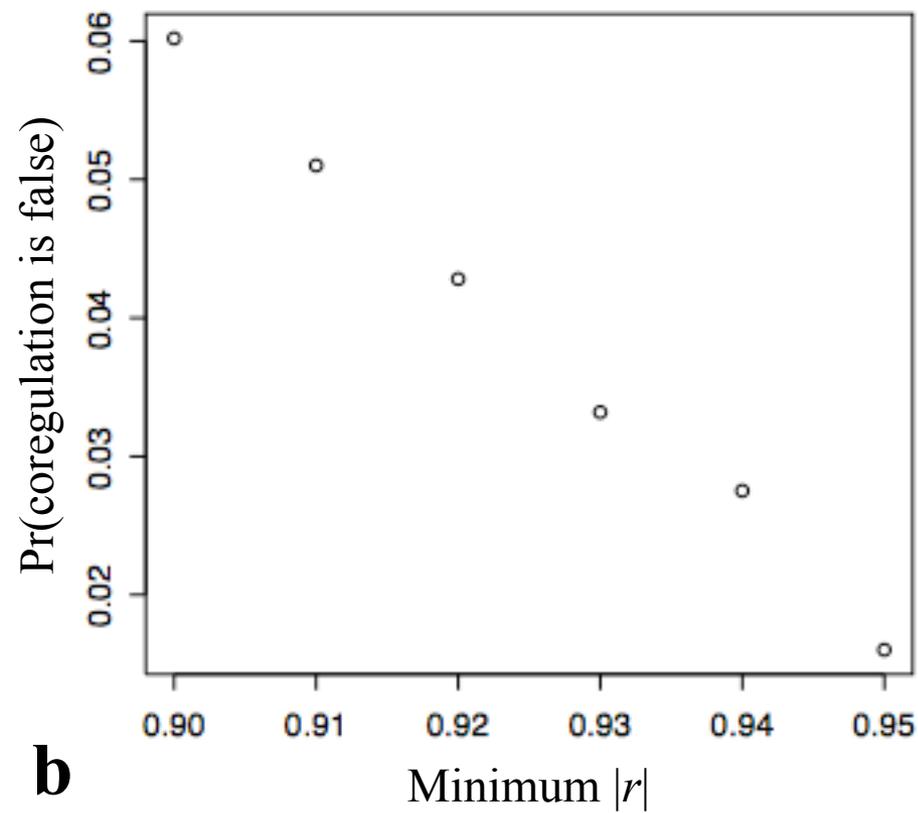

a    b

# Figure 3

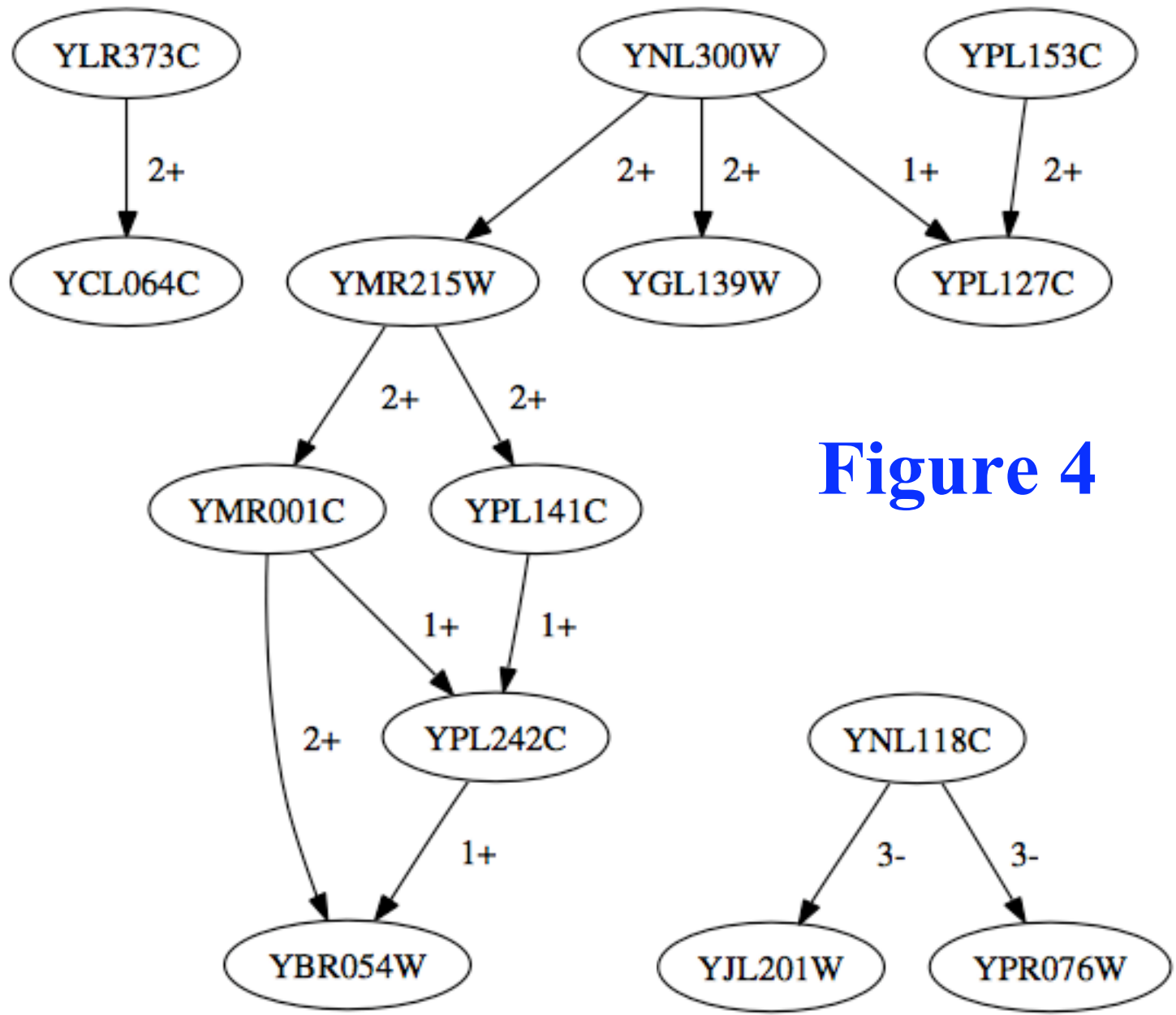

# Figure 5

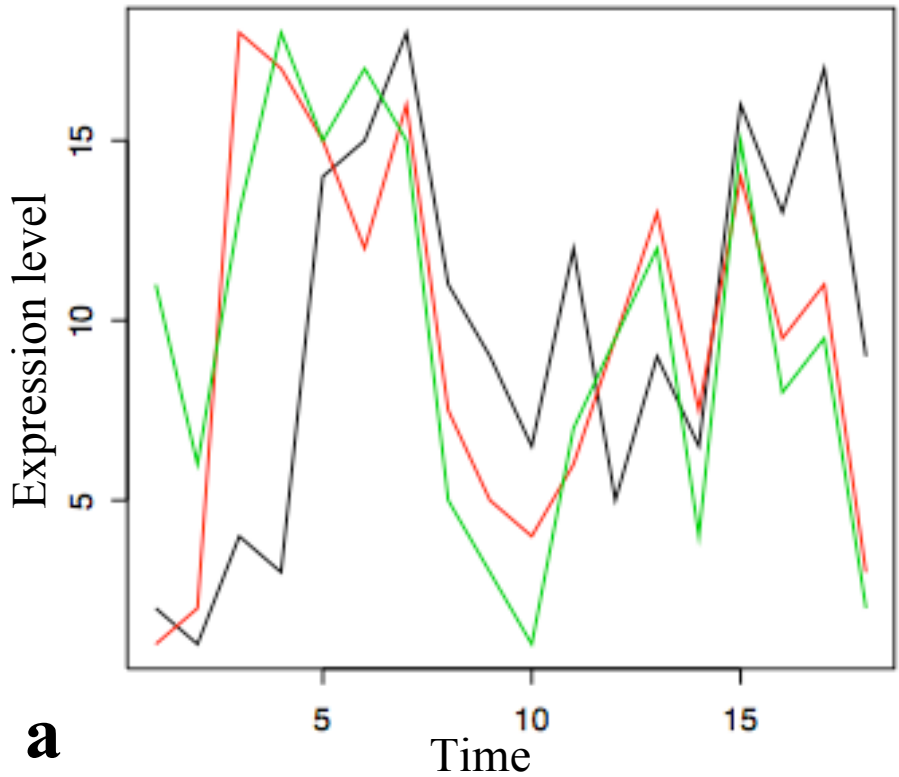 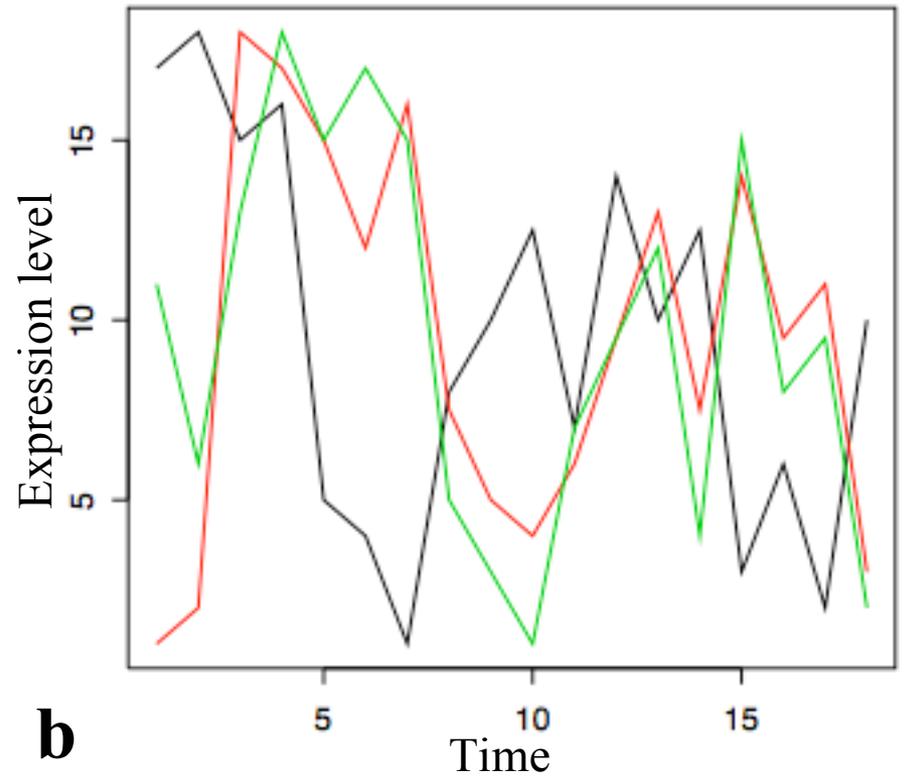

**a** Three genes  **b** …leading gene reflected

# Figure 6

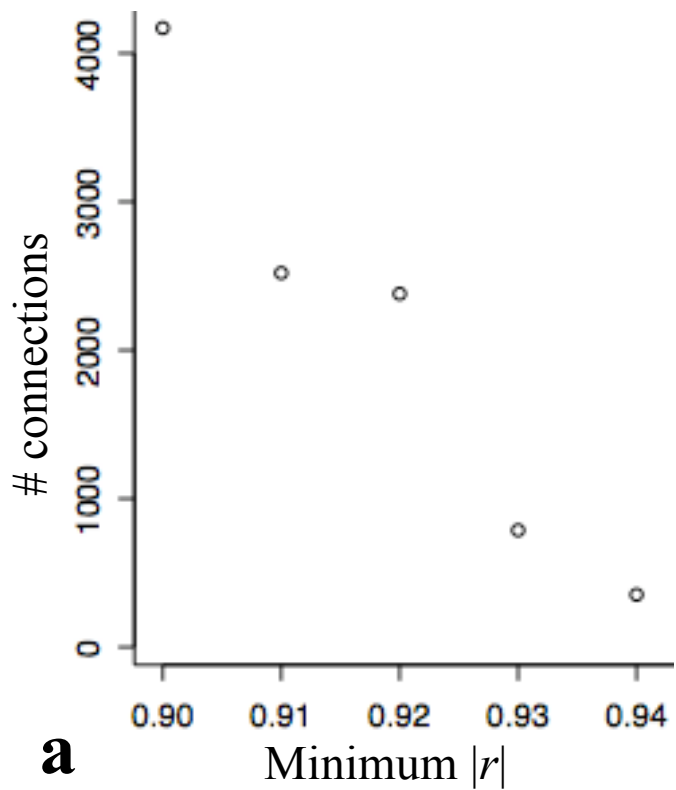 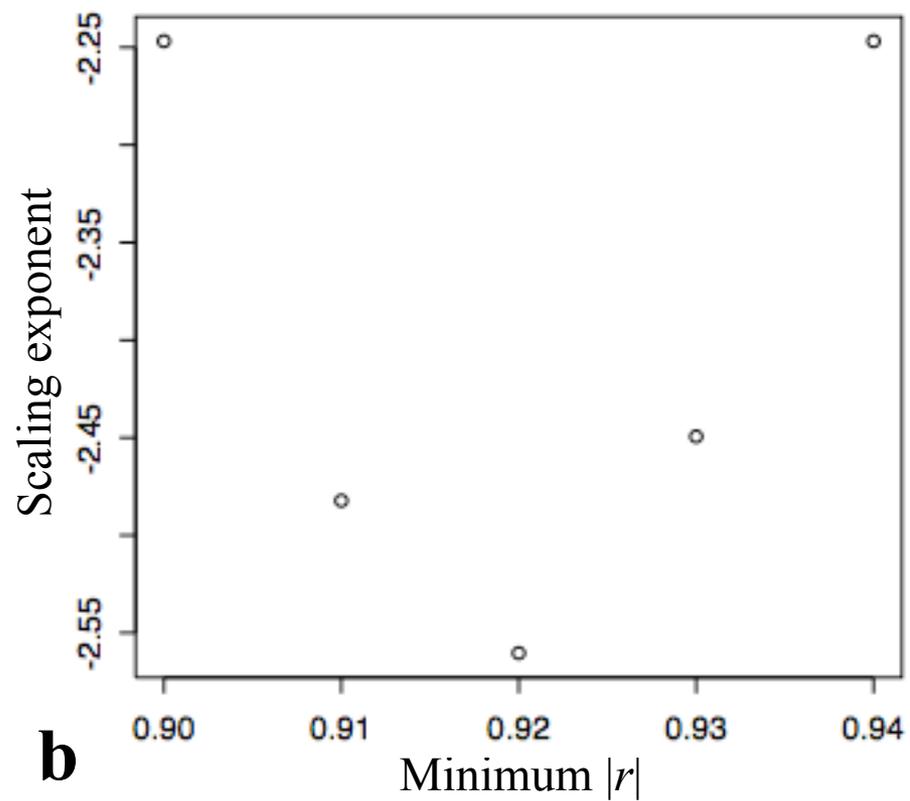

a    b

# Figure 7

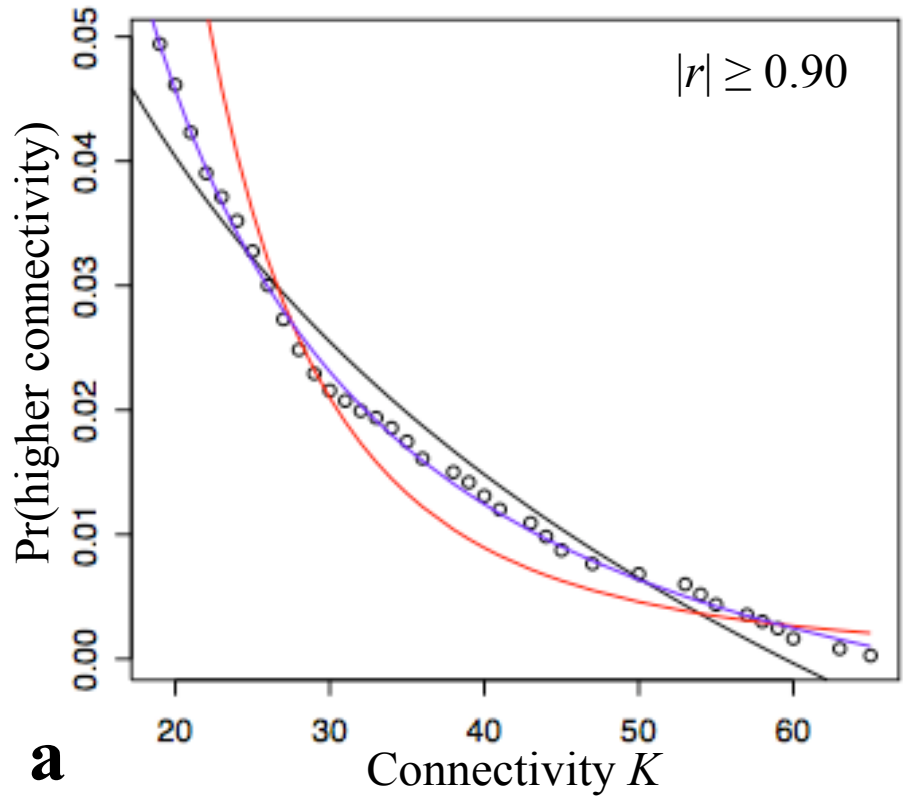 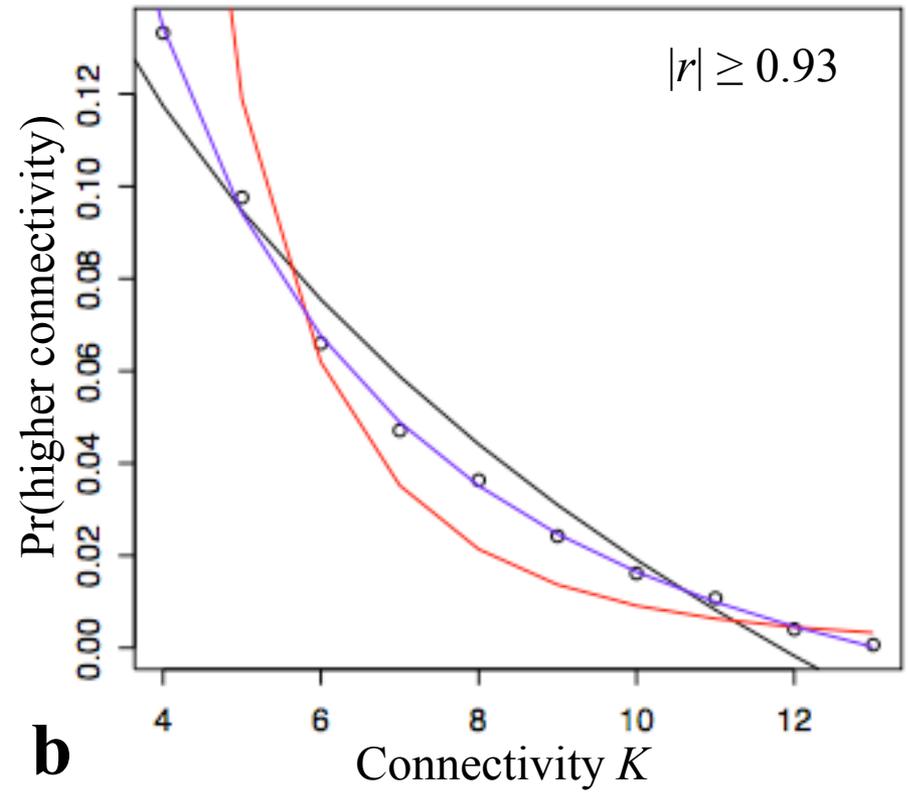

a  b